\documentclass{aa}
\usepackage{natbib}
\usepackage{graphicx}

\usepackage{txfonts}
\newcommand{\ve}[1]{\mathbf{q1}}

\newcommand{\f}{\frac}

\newcommand{\be}{\begin{equation}}      
\newcommand{\ee}{\end{equation}}      
      
\newcommand{\bef}{\begin{figure}}      
\newcommand{\eef}{\end{figure}}      
\newcommand{\bea}{\begin{eqnarray}}    
\newcommand{\eea}{\end{eqnarray}}

\newcommand{\av}[1]{\ensuremath{\left\langle q1 \right\rangle}}

\newcommand{\tve}[1]{\tilde{\boldsymbol{q1}}}

\def\bse{\begin{subequations}}
\def\ese{\end{subequations}}

\def\lsim{\raise 0.4ex\hbox{$<$}\kern -0.8em\lower 0.62ex\hbox{$\sim$}} 
\def\gsim{\raise 0.4ex\hbox{$>$}\kern -0.7em\lower 0.62ex\hbox{$\sim$}}

\def\f0N{f_0^{(N)}}
\def\bec{\begin{center}}
\def\eec{\end{center}}

\begin{document}

\title{Gravitational collapse from cold uniform asymmetric initial conditions}
  
\titlerunning{Gravitational collapse from cold uniform asymmetric initial conditions}
  
\authorrunning{Sylos Labini \& Joyce}  
  
  \author {Francesco  Sylos Labini \inst{1,2} and Michael Joyce \inst{3}}

        \institute{
          Centro  Ricerche Enrico Fermi, Via Panisperna 89 A,
           00184 Rome, Italy 
          \and 
          Istituto Nazionale Fisica Nucleare, Unit\`a Roma 1, Dipartimento di
          Fisica, Universit\`a di Roma ``Sapienza'', Piazzale Aldo
          Moro 2, 00185 Roma, Italia 
          \and 
          Laboratoire de Physique Nucl\'aire et de Hautes \'Energies, UPMC IN2P3 CNRS UMR 7585,
          Sorbonne Universit\'e, 4, place Jussieu, 75252 Paris Cedex 05, France}

\date{Received / Accepted}

\abstract{
Using controlled numerical N-body experiments, we show how,
 in the collapse dynamics of an initially cold and uniform 
distribution of particles with a generic asymmetric shape,  finite $N$ fluctuations
and perturbations induced by the anisotropic gravitational field compete to  determine 
the physical properties of the asymptotic quasi-stationary state. When finite $N$  fluctuations 
dominate the dynamics, the particle energy distribution changes greatly and the final density profile 
{decays outside its core} as $r^{-4}$ with an $N$-dependent amplitude. 
On  the other hand, in the limit where the anisotropic perturbations dominate,  the collapse 
is softer and the density profile shows a decay as  $r^{-3}$, 
as is typical of  halos in cosmological simulations. However, even in this limit, convergence with $N$ of the 
macroscopic properties of the virialized system,
such as the particle energy distributions, the bound mass, and the density profile, is very slow and not clearly 
established, including for our largest simulations (with $N \sim 10^6$). Our results illustrate the challenges of 
accurately simulating the first collapsing structures in standard-type cosmological models.}

\maketitle
\keywords{ methods: numerical; galaxies: elliptical and lenticular, cD; galaxies: formation}

\section{Introduction}
The evolution under Newtonian self-gravity of an isolated distribution of particles that start from initial conditions
(ICs) in which the particles are initially at rest and are randomly distributed in a sphere with uniform mean density has been studied by numerous authors since the earliest numerical simulations of such systems. Numerous variants have also been explored, notably with initial velocity dispersion (using radially dependent density profiles) and for spheroidal boundaries 
\citep{Henon_1964,henon_1973,vanalbada_1982,merritt+aguilar_1985,aarseth_etal_1988,aguilar+merritt_1990,theis+spurzem_1999,   boily_etal_2002, roy+perez_2004, boily+athanassoula_2006, barnes_etal_2009,joyce_etal_2009, syloslabini_2012, syloslabini_2013,worrakitpoonpon_2014, SylosLabini+Benhaiem+Joyce_2015, Benhaiem+SylosLabini_2015,Benhaiem_etal_2016, Benhaiem+SylosLabini_2017,SpeRCD2017,SylosLabini_etal_2020,Worrakitpoonpon_2020}. 
Such systems provide a natural and simple setting to explore the rich complexity of the nonlinear phase of gravitational dynamics relevant to many different problems in astrophysics and  cosmology. 
Though in the context of structure formation in cosmology structures are never strictly isolated, these simple models are nonetheless highly relevant.
Indeed, the so-called spherical collapse model, which analytically describes
 a smooth spherical over-density in an expanding universe,
is a paradigmatic model that guides understanding of the nonlinear cosmological structure formation (see, e.g., \citealt{Peebles_1983}). Despite 
its simplicity, and indeed its breakdown in a singularity at a finite time, it describes qualitatively well what is actually observed 
in $N$-body simulations of standard cosmological models: Over-dense regions expand more slowly on average until  
they  ``break off'' from the expansion and undergo a strong collapse, starting from conditions that are effectively those of an 
isolated, initially cold configuration. The specific design of any set of numerical experiments in either the astrophysical or cosmological
context varies depending on the motivation of the authors and the physical issues they focus on. We report here a study of a very
simple variant of this class of models. It is designed to explore how deviations from spherical symmetry compete with the density 
fluctuations intrinsic to the discrete particle distribution in regulating the singularity in the limit of a cold uniform collapse
(i.e., of the spherical collapse model in cosmology). The understanding of the dynamics that can result from the very 
simple class of ICs we consider -- with just one parameter in addition to the particle number -- is of broad 
interest. More specifically, as we explain,  it gives insight into the challenges of accurately simulating the first 
nonlinear structures in many cosmological models.

One particular question that can be addressed in a controlled manner with these systems is the extent to which the evolution of 
discrete  $N$-body systems approximate, and converge to,  
in an appropriate $N \rightarrow \infty$ limit,  the continuum limit given by the 
Vlasov-Poisson system (or collisionless Boltzmann equations) for a smooth phase-space density. 
Indeed, for the relevant physical problems, the particles of the $N$-body simulations 
are usually thought of not as physical particles but simply as 
a tool for representing the physical phase-space density, the evolution of which 
is described by these equations. 
In the cosmological context the physical particles are usually microscopic dark matter particles, 
while the particles of simulations are unphysical ``macro-particles'' with astrophysical-scale masses. An important question is 
then to what extent the results obtained from these simulations are dependent on $N$ and what the associated limit on their
accuracy is. In more physical terms the question can be formulated as to what extent the fluctuations associated
with the particle discretization play a role in the dynamics in its nonlinear phase. 
This question of $N$-body discretization effects in cosmology
can be straightforwardly addressed using isolated systems because, as noted, we can consider, in light of the exact spherical collapse model, that nonlinear 
structure formation can be approximated by the collapse of quasi-isolated initially cold clumps.

In current models of cosmological
structure formation, there is always a minimal scale in the initial spectrum of fluctuations, the physical size of which varies greatly depending on whether the supposed dark matter is ``cold'' or ``warm'' (see, e.g., \citealt{Wang_White_2007}, \citealt{KUHLEN201250}, and references therein).
The first clumps
therefore form starting from fluctuations at this scale and are initially highly uniform at smaller scales. In the discretized version
of the problem (i.e., as represented in an $N$-body simulation) there are, in addition, the fluctuations associated with the
particle discretization. The question of discreteness effects thus appears to be well and simply posed  in terms of the  $N$ dependence 
of the evolution of an isolated, initially cold, uniform clump.  However, things  
are not  so simple because, in spherical
symmetry, the $N \rightarrow \infty$ limit has the finite time collapse singularity of the spherical collapse model. 
Spheroidal clumps are a simple alternative, but they are also characterized by a singularity in the same limit due to
their residual rotational symmetry \citep{Lin_Mestel_Shu_1965,Worrakitpoonpon_2020}. 

In a recent article,
 \cite{SylosLabini_CapuzzoDolcetta_2020} considered ICs that are spherical but with 
a simple family of density fluctuations within the sphere of which the amplitude is regulated by a single 
additional parameter. This reproduces ICs that are expected to model clumps forming from
preexisting structures at smaller scales, as at later times in typical cosmological models. Here we 
instead consider modeling the clumps as being uniform at small scales but having an irregular shape. This is 
a good schematic representation of the first collapsing structures in a standard cosmological model.

The cold uniform spherical case, with a single free parameter, $N$,  has been studied at length
in, for example, \cite{aarseth_etal_1988} and \cite{joyce_etal_2009}. 
In this case the singularity in the $N \rightarrow \infty$ is 
thus regularized by the density fluctuations associated with discreteness alone. More specifically,
it is found, for Poisson-distributed particles, that the minimal system radius 
approximately scales  as $r_{min} \sim N^{-1/3}$. This behavior can be derived from  
the growth of the density perturbations  in the bulk, which evolve until the associated peculiar kinetic energy 
becomes on the order of the gravitational potential energy. 
Thus, the smaller $N$ is (i.e., the larger the initial density fluctuations), the 
softer the collapse is: As a consequence of the larger resulting particle peculiar velocities, 
the collapse phase stops earlier. 
In addition, the  properties of the virialized 
quasi-stationary state (QSS; e.g.,  the amplitude of the density profile and the particle energy distribution) that is 
attained rapidly after the collapse are $N$ dependent. In the model considered 
by  \cite{SylosLabini_CapuzzoDolcetta_2020}, the  particles are instead distributed in $N_c$ spherical sub-clumps of radius $r$; $r$ is smaller than the radius of the system itself, $R$,
which allows the  amplitude of initial fluctuations to be regulated. 
As their amplitude is increased,
the very violent collapse of the simple model is progressively ``softened'': 
While in the former 
case there is a large change in the particle energy distribution driven by a strong variation in 
the system's mean field,  in the latter there is a marginal change in the system's phase-space distribution.  
The properties of the QSS reached thus depend strongly on whether the collapse was 
violent or soft:  In particular, the final density profile decreases as $\sim r^{-4}$  when it is very
violent, while it decreases as $\sim r^{-3}$ in the soft limit. In the model we explore here, we again see these same limiting behaviors emerge.

The term ``violent relaxation'' that we employ here was introduced by
\cite{lyndenbell}  to describe the collective mean-field relaxation
process that drives a self-gravitating system prepared in an out-of-equilibrium IC
to form a QSS. Lynden-Bell also proposed a theory to calculate the resulting QSS
using a statistical approach based on the collisionless Boltzmann equation.
The predictions of this theory have, however,  been shown by 
the results of numerical experiments 
(see, e.g., \citealt{arad+lyndenbell_2005}) to not account well for the relaxed QSS of such systems.
We note that, for the ICs we study here,  the theory of
Lynden-Bell cannot be adequate for a very simple reason: It assumes
the mass and energy of the QSS to be that of the initial state, while these
systems are always characterized -- because of the violence of the relaxation --
by the ejection of a very significant fraction of their initial mass and energy.
We note that a very different semi-analytic model of violent relaxation that
does admit the presence of ejected mass has been proposed by 
\cite{Mangalam_etal_1999}. It is based on the physical picture of strong 
and short-lived interactions between the system's potential fluctuations 
and particle orbits and, interestingly, also accounts for an associated
QSS density profile decreasing as $\sim r^{-4}$.

In this work we  introduce and use numerical experiments to study a simple toy model in which
the perturbations of shape and the finite $N$
fluctuations can be varied independently by tuning two parameters: 
a ``shape parameter,'' $p$, which controls the degree of the initial spherical 
symmetry breaking,  and the particle number, $N$, which controls 
the amplitude of discrete density fluctuations.
These latter fluctuations are chosen from two different types:
uncorrelated Poisson fluctuations and those in a  
highly uniform ``glassy'' distribution 
This model thus allows us to study, in a controlled way, the competition between 
finite $N$ fluctuations and  anisotropic perturbations. 
We show that the evolution, and thus the properties of the QSS formed 
after the collapse, crucially depends on which of these two competing 
effects dominates the dynamics. In addition, we show that even when 
shape perturbations are predominant, 
the convergence as a function of $N$ is very slow.

The paper is organized as follows. In Sect. \ref{InitialConditions}
we describe the generation of ICs 
and give some details about the numerical simulations. 
In the following section we present our results. Finally,
in Sect. \ref{Conclusions} we draw our conclusions, in particular for what 
our study tells us about resolution issues related to discretization in cosmological 
$N$-body simulations.

\section{Initial conditions}
\label{InitialConditions}

\begin{figure} 
\includegraphics[width=3.8in,height=4.5in,angle=-90]{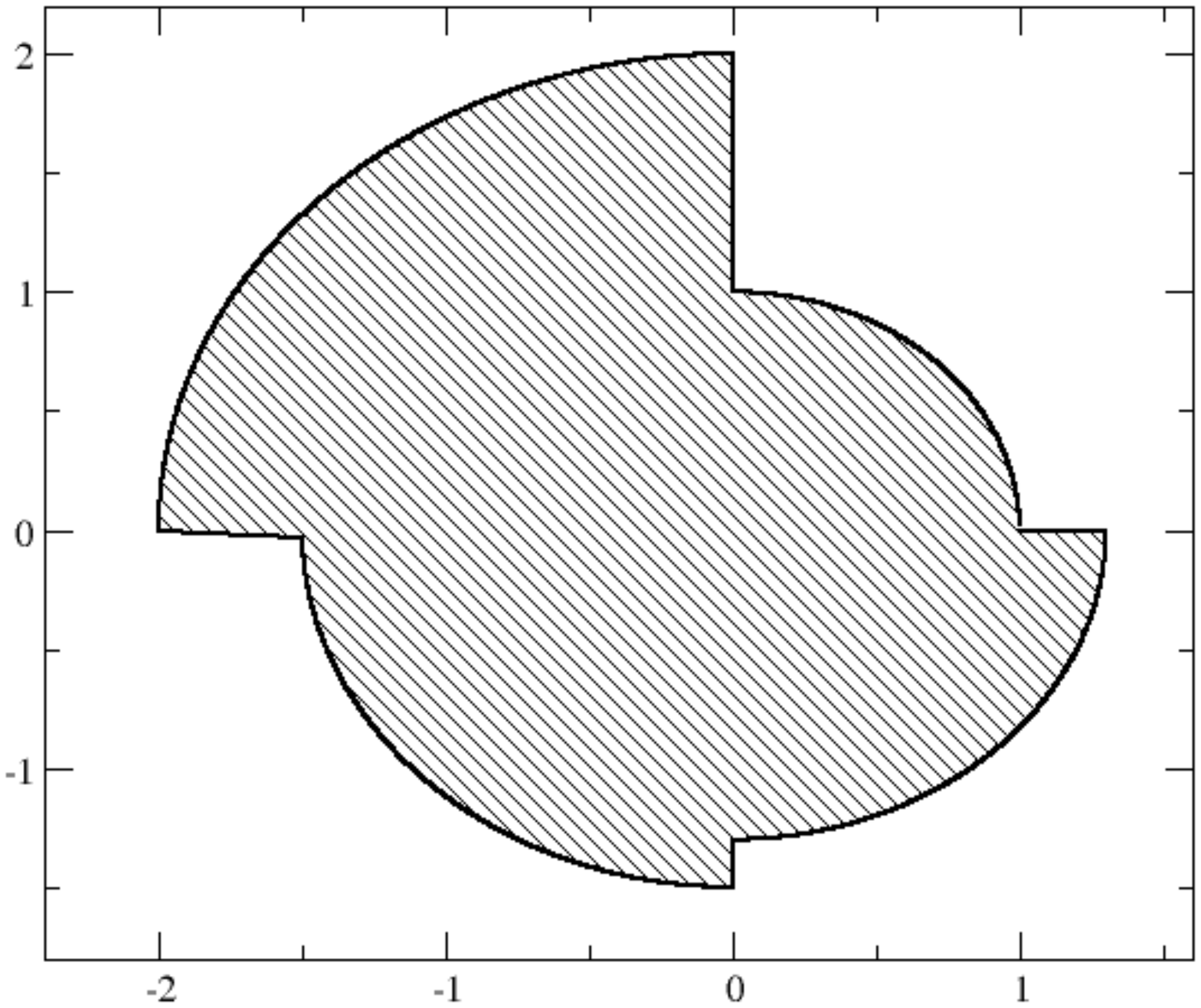}
\caption{Schematic representation (2D projection) of our ICs for $p=1$. 
}
\label{S5P10_xy_t0} 
\end{figure}

The ICs for our numerical experiments consisted of 
particles distributed with uniform mean density in a connected region 
with an asymmetric shape characterized by a single dimensionless
parameter, $p$.  These ICs were generated as follows. 
The region was defined by a sphere of radius $R_0$ (hereafter ``the inner sphere''), 
to which a spherical shell divided into eight identical 
contiguous sectors of solid angle $\Omega=\pi/2$ was added. The outer
radius of each shell was set equal to $R_i=R_0\times (1+p_i),$ where $p_i$ is a 
random number with uniform distribution in the range $[0,p]$
and the index $i=1,...,8$ runs over the sectors (see Fig. \ref{S5P10_xy_t0}).  
The particle configurations were then generated inside this region
in one of two different ways. For the ``Poisson'' configurations,
we first randomly distributed (without correlation of position) 
$N_0$ particles in the inner sphere such that their mean density in
this region is 
\be
n_0 = \frac{3N_0}{4 \pi R_0^3} \;.
\ee
We then distributed in the $i^{th}$ sector ( for $i=1,...,8$) 
the number of particles that gives the same mean 
number density in each  sector, namely,
\be
N_i= \frac{\pi R_i^3 n_0}{8} \;.
\ee
The total number of particles is thus $N=\sum_{i}^8 N_i$.
For the ``glass'' configurations we extracted a region 
with the appropriate shape and size from a large glass, a very 
uniform configuration obtained by running an N-body simulation
with a repulsive $r^{-2}$ force in periodic boundary conditions 
until an equilibrium configuration is obtained, following
a procedure often employed in the generation
of ``pre-initial" conditions that represent the unperturbed
universe for cosmological simulations \citep{springel_2005}.

The limit  $p=0$ corresponds to spherical symmetry. For $p\neq 0$, all rotational symmetry is broken, 
and the larger $p$ is the more asymmetrical the distribution. We considered here only the range $0\leq p \leq 1$.
For  $p>1$, the evolution typically leads to fragmentation into multiple 
substructures, which then, on a longer timescale, merge together. In this case, 
the dynamics is thus qualitatively different. In this work 
we explored up to $p=1,$ for which the dynamics can be 
described as  a single or a ``monolithic'' collapse.

We ran simulations for a ``grid'' of ICs in our 
two-parameter space: for each of $p=0,0.2,0.4,0.6,0.8,$ and $1$.
The particles were of equal mass and 
we took $N_0$ in the range $10^3-10^6$.
To designate results in the simulations, we used, for
given specified $p$, a label  consisting of a letter and an integer. The latter is simply 
$\log_{10} N_0$, while the letter is either $S$ or
$G$: $S$ means that the particles are in 
a Poisson configuration, and $G$ means that they
are in a glass configuration.
For example, $S4$ indicates that the particles were initially Poisson distributed 
and that $N_0=10^4$. We note that we thus label our simulations by $N_0$, 
the number of particles inside the inner sphere, and not by the (slightly different)
total number of particles. This is appropriate because it allows us to compare
simulations at a fixed (or varying) particle density. 

Hereafter, distances are given in units of the initial radius  of the inner spherical system, $R_0$, 
and time is given in units of 
 \be
 \label{tau0}
 \tau_0 = \sqrt{ \frac{\pi^2 R_0^3}{8 G M_0}} \;, 
 \ee
where $M_0$ is the mass of the inner spherical system and $ \tau_0$ is the characteristic time
corresponding to the time at which the maximal contraction of the system is attained. Virialization
occurs a short time later, and in all our simulations the system is well relaxed to 
a QSS before $t=2$ in these units. 

We report results for the density profile, $\rho(r),$ normalized by the total mass. 
Further, we calculate
\be
\label{eq:delta}
\Delta (t) = \frac{1}{\langle e(0) \rangle } 
\sqrt{ \frac{\sum_{i=1}^N (e_i(t) - e_i(0))^2} {N} } 
,\ee
{where $e_i(t)$ is the energy of particle $i$ at time $t$ and}
\be
\langle e(0) \rangle =   \left( \frac{\sum_{i=1}^N e_i(0)}{N} \right) \;
\ee
is its initial average value. The $\Delta (t)$ thus gives a 
simple measure of the total change of the particle energies relative to the initial time 
in units of the initial average energy per particle. 
When $\Delta(t)  > 1$, the energy change is large and the collapse is violent, 
while   when $\Delta(t)  <1$  the energy change is small and the collapse is 
soft \citep{SylosLabini_CapuzzoDolcetta_2020}.

Numerical simulations were run using the code Gadget \citep{springel_2005}. We refer to \cite{SylosLabini_CapuzzoDolcetta_2020}  for further discussion 
about the details of the  numerical integration: Here we just specify that the gravitational softening length 
was taken to be $1/100$ of the minimal radius, $r_{min}$, reached by the system  during the collapse. 
We performed numerous tests with smaller and larger softening to check the convergence,
 finding that 
 results do not depend significantly on $\varepsilon$  as long as $\varepsilon < r_{min}$.  
Total energy is conserved to a precision better than a few percent; we tested 
that this is enough to have good convergence of the macroscopic quantities measured here.

\section{Results} 
\label{Simulations}

\subsection{Particle energy distributions}
 
The temporal  evolution of the  particle energy distribution, $P(e,t)$,
is a key statistical quantity that characterizes the 
collapse dynamics \citep{SylosLabini_CapuzzoDolcetta_2020}. 
Figure \ref{P_Pe_S5}  shows its behavior for the different $p$ simulated
for Poisson-distributed particles and $N_0=10^5$ at 
$t=0$ and $t=9$ (when, as noted above, the system is very
well virialized).

The trend, which can be seen in Fig. \ref{P_Pe_S5}, is very clear:  
In all cases the virialized energy distribution is very different to,
and in particular very much broader than,  the initial one.
However, this difference is greatest at $p=0$ and becomes
progressively smaller as $p$ increases. This is readily explained by
 considering that $p$ controls the 
violence of the collapse: When $p=0$, the collapse is close to
homologous {and the singularity of the  $N \rightarrow \infty$ limit is regulated only 
by the density fluctuations seeded by  the initial discretization.
On the other hand,  for {$p \neq 0$} the collapse is no longer singular in this same 
limit because tidal effects induced by spatial anisotropies  change the collapse times 
of shells that are initially located 
at different distances from the center and/or at different angular locations.  
The larger $p$ is, the more important the effect of these anisotropies in
modifying the collapse  is and the smaller the change of $P(e,t=9)$ with 
respect to $P(e,t=0)$. 
One can also see in the plots that there is a small additional peak 
around $e=0$ that appears at $p=0.2$ and becomes ever more apparent
as $p$ increases. Further, its position (in energy) clearly  coincides 
more closely with the center of the initial distribution as 
$p$ increases. This feature is evidently a residual of the IC and
gives a further quantitative measure of how violent the collapse is. 
\begin{figure*} 
\includegraphics[width=5.0in,angle=-90]{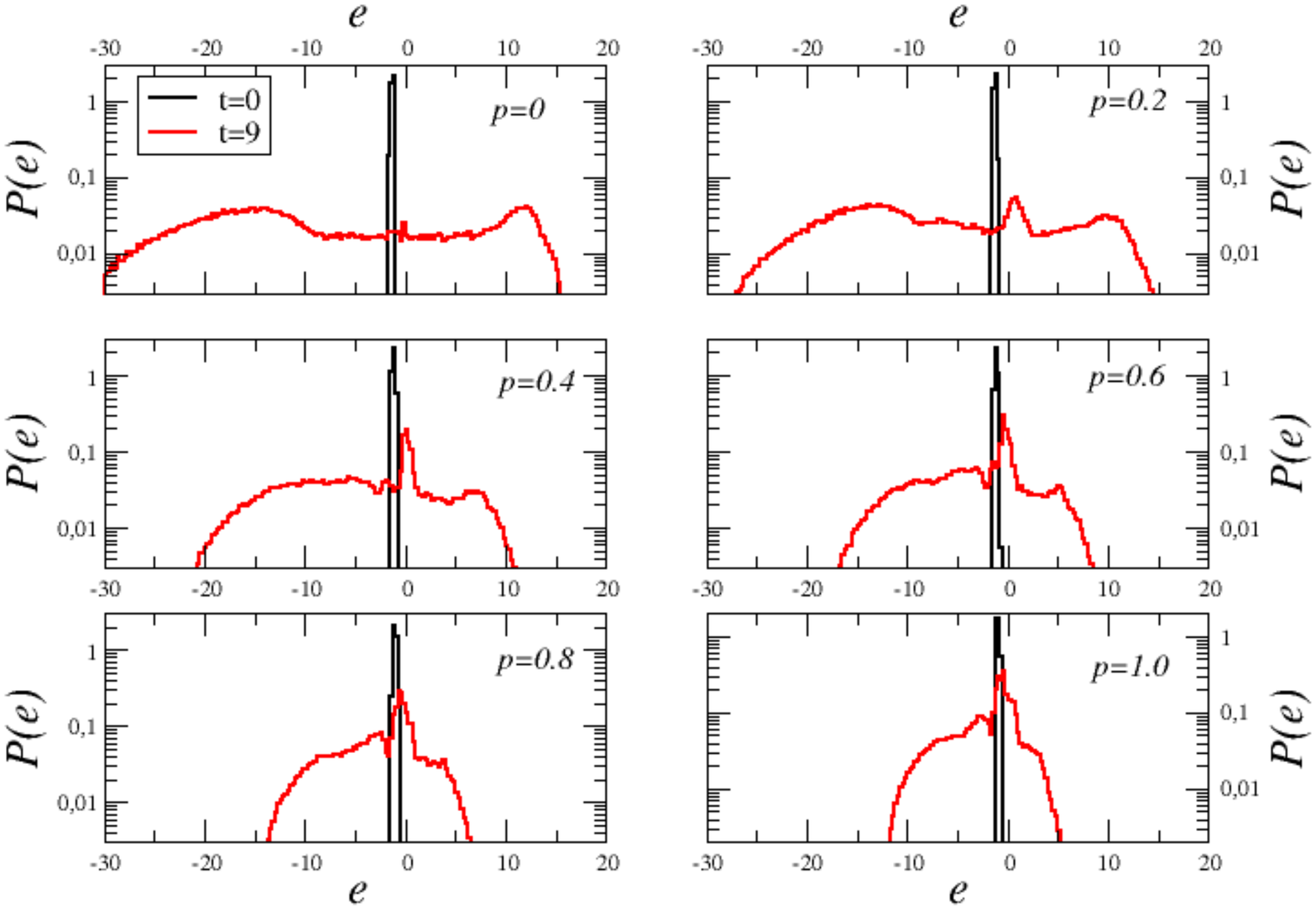}
\caption{Energy distribution at $t=0$ (black line) and at $t=9$ (red line) for the series S5; 
each panel shows a different value of the initial breaking of spherical symmetry, $p$.}
\label{P_Pe_S5} 
\end{figure*}

Next we considered how the $N$ dependence of the energy distributions is obtained
from the different ICs.
Figure \ref{P_Pe} shows  $P(e,t)$, again at $t=9$,  for  $p=0$ (top panel) and $p=1$ (bottom panel)
for the different indicated values of $N_0$ and initial Poisson configurations.
We observe that while for $p=0$ the energy 
distribution $P(e,t=9)$  depends strongly on $N$, 
for $p =1$ there are only very much smaller residual finite $N$ effects. 
Nevertheless, in the tails of the distribution for $p=1$
-- for both the most bound and most energetic unbound particles  --
there appears to still be some systematic dependence on $N$, with the tails 
becoming marginally more spread for larger values of $N$.
The convergence to an $N$-independent result thus  seems to be slow,
with visible evolution with $N$ still at $N_0=10^6$. 
\begin{figure} 
\includegraphics[width=2.8in,angle=-90]{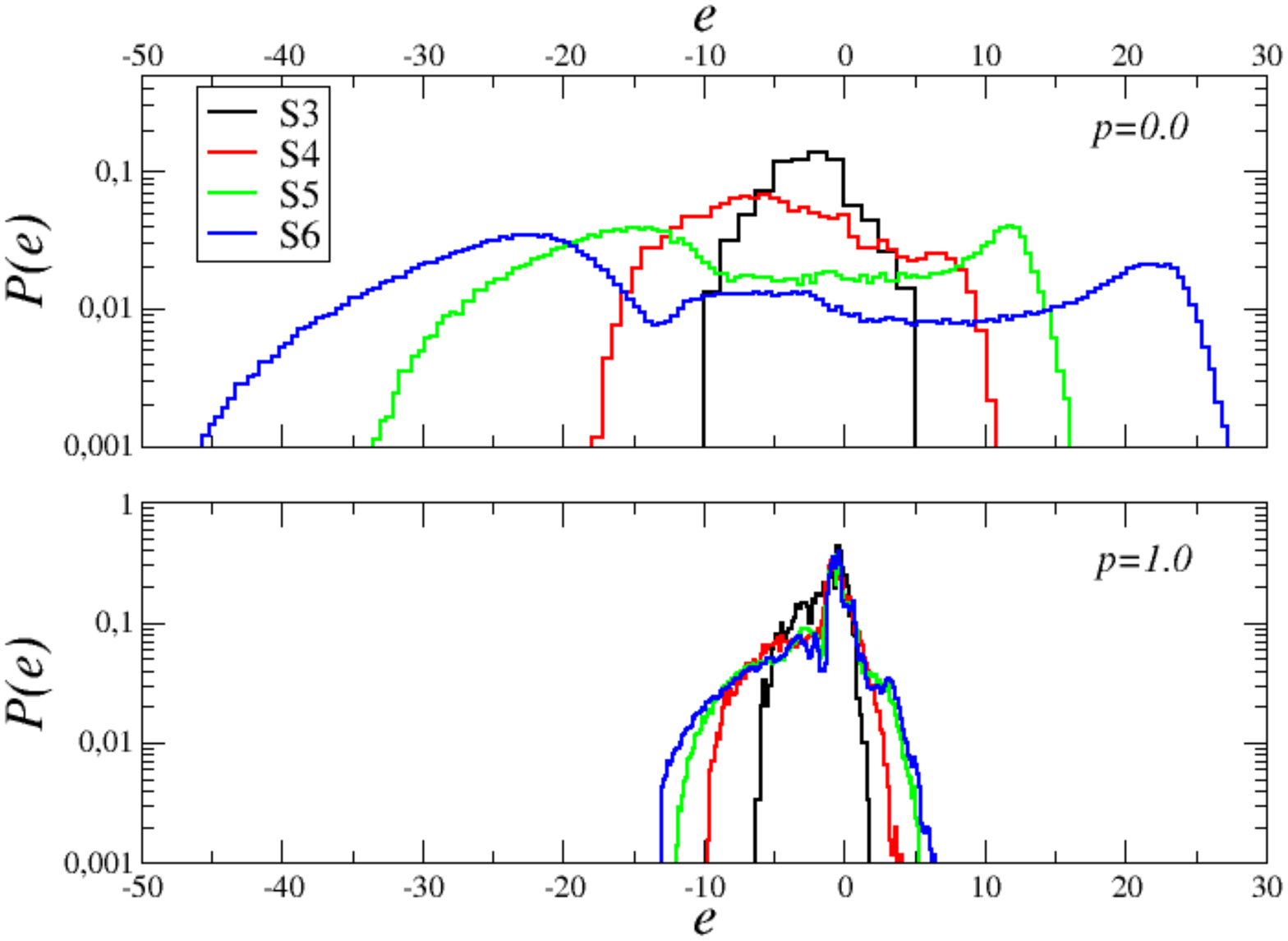}
\caption{Energy distribution at $t=9$ for $p=0$ (upper panel) and
$p=1$ (bottom panel) for the four realizations with a different 
number of Poisson-distributed particles, i.e., $N_0=10^3, 10^4, 10^5$, and $10^6$.}
\label{P_Pe} 
\end{figure}

\subsection{Mean energy exchange}

The upper panel of Fig. \ref{Delta_P} shows the behavior of $\Delta(t=9)$, as defined
above in Eq. \ref{eq:delta}, as a function of $p$ for initial Poisson configurations
with different $N_0$. As for $P(e,t)$, the choice of the time $t=9$ is somewhat arbitrary
as this quantity is essentially time independent provided $t  \ge 2$. 
We observe that in all cases  $\Delta>1$, which implies
that the collapse is always quite violent in the sense that the  
variation in  the particle energy distribution is large.
In  a collapse from cold ICs,  $\Delta<1$ can be  obtained 
in the presence of initial fluctuations that are sufficiently large
such that  clustering proceeds in a bottom-up manner \citep{SylosLabini_CapuzzoDolcetta_2020}. 
We observe that $\Delta(t=9)$ decreases monotonically as
a function of $p$ at a fixed value of $N_0$.
The behavior simply reflects, once again, that the collapse 
``softens'' progressively as $p$ increases. On the other
hand,  $\Delta(t=9)$ also increases with $N_0$ at any
fixed $p$, which indicates that the collapse is, 
in all cases, becoming more violent as $N$ increases.
There is, however,  a clear weakening of this $N$ dependence 
as  $p$ increases and an apparent onset of good 
convergence for $p=1$  with the two runs with the 
largest particle numbers close together. 
Thus, it appears that at $p=1$ we attain,
at $N_0 \sim 10^5$,   
the limit where the collapse is regulated by the shape fluctuations alone.

\bef
\includegraphics[width=2.8in,angle=-90]{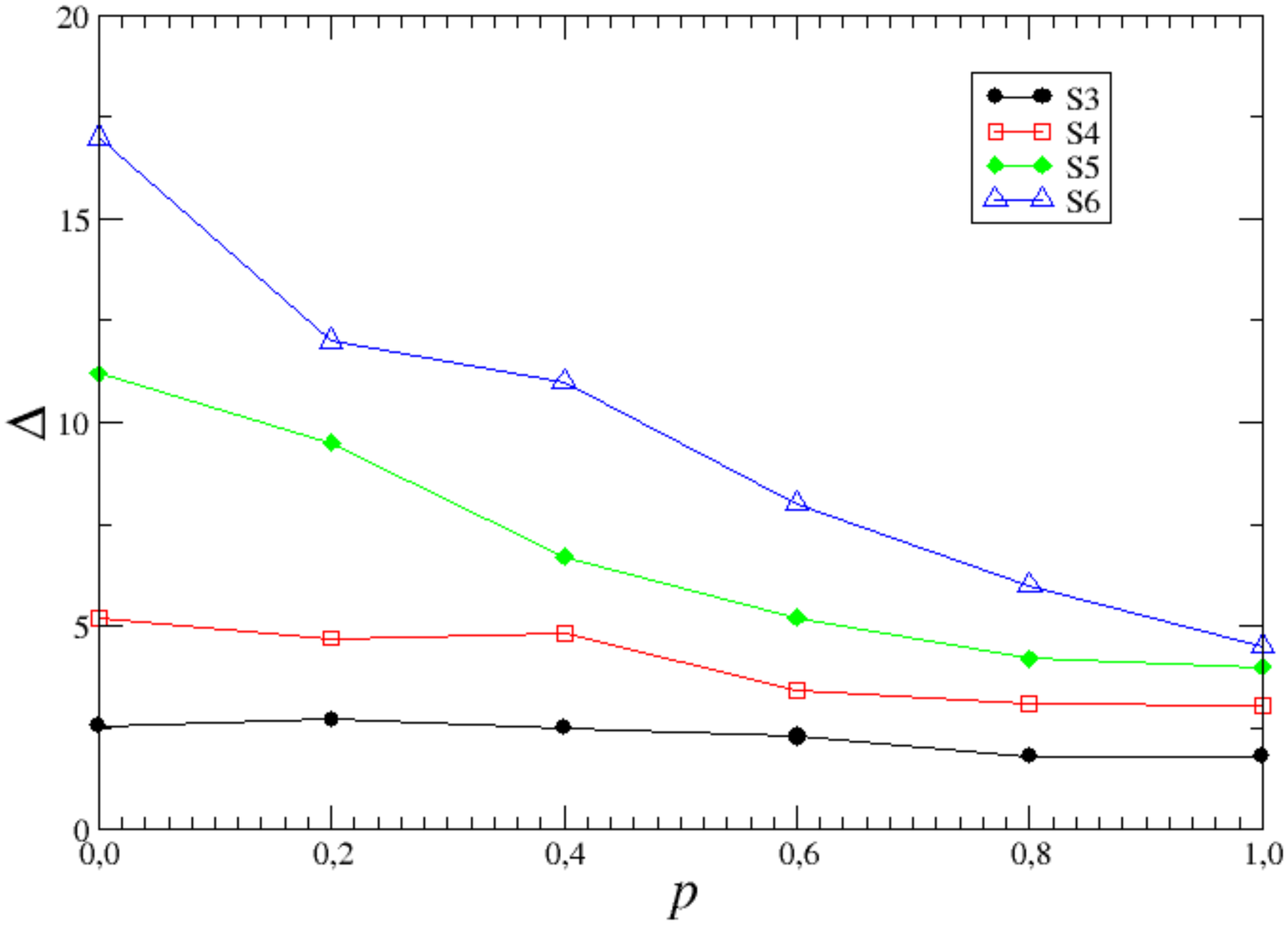}
\includegraphics[width=2.8in,angle=-90]{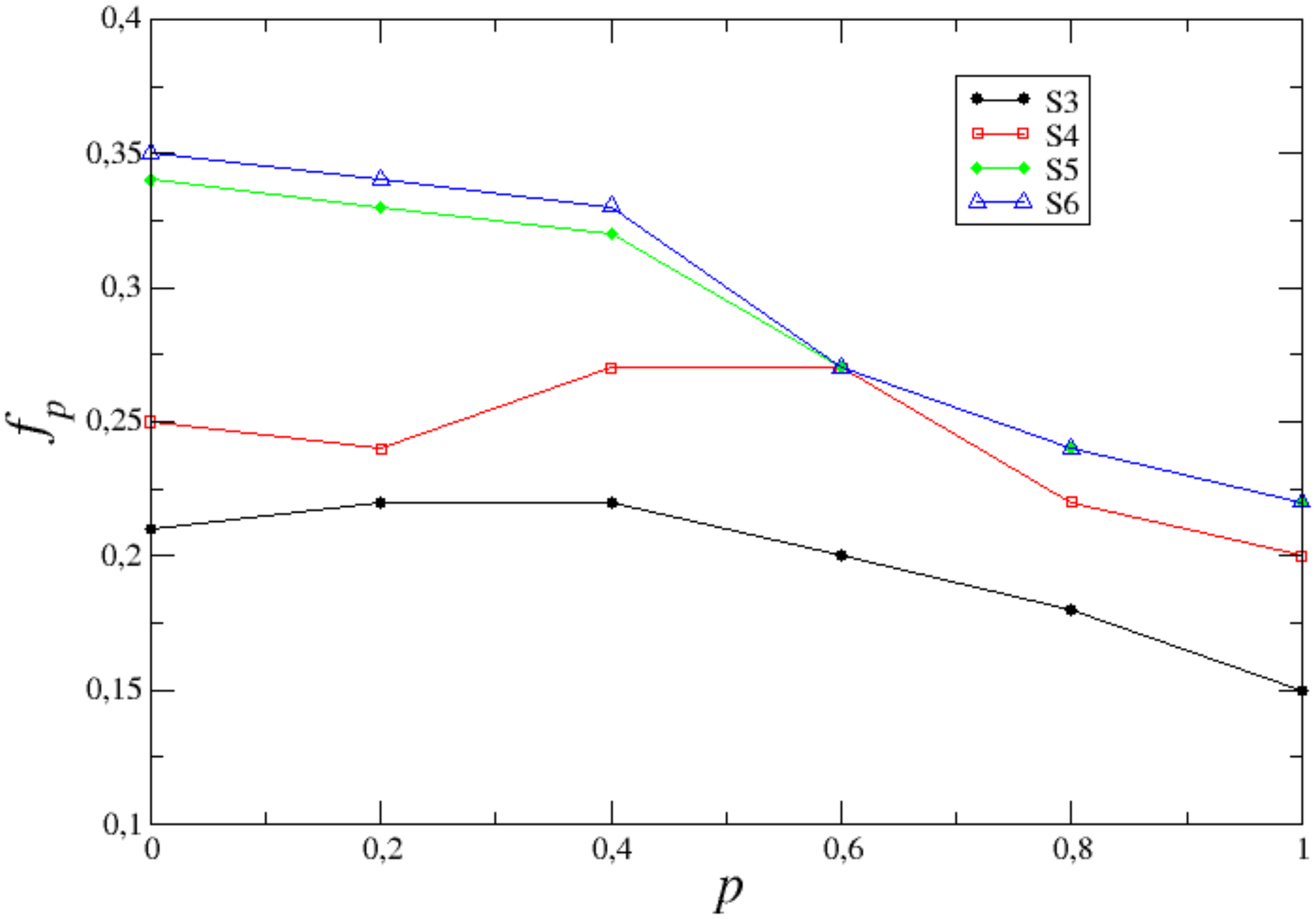}
\caption{
Quantifying the violence of the collapse. 
Upper panel: Energy exchange $\Delta$  (see  Eq. \ref{eq:delta}) at time $t=9$ 
 as a  function of $p$ (with different $N_0$).  
Bottom panel: Same but the plotted quantity $f_p$ is the fraction of ejected particles, i.e., particles 
that have positive energy after the collapse}.
\label{Delta_P}
\eef

\subsection{Ejected mass}

The bottom panel of  Fig. \ref{Delta_P}  shows the 
fraction of ejected particles, $f_p$, as a function of the initial asymmetry, $p$,  for simulations with $N_0=10^5$. As described in detail in \cite{joyce_etal_2009},
 \cite{syloslabini_2012}, and \cite{SylosLabini_CapuzzoDolcetta_2020},
 these are essentially all particles whose energy becomes
positive in a very short interval around the time of maximal contraction and remains positive thereafter.
We note the very large values (approximately one-third of the mass) attained for the spherical IC
(see \cite{joyce_etal_2009} for further detail) characterized by the most violent collapse. The behaviors seen
are very similar to those in the previous plot, except that there appears to be clearer evidence for
convergence with $N$  at the larger values  of $p$. 

\subsection{Density profiles}

Figure \ref{P_rho} shows normalized density profiles, $\rho(r)$, at $t=9$ for the indicated simulations.
We focus on the very well resolved regions, where $r$ is significantly larger than the gravitational
softening. We have multiplied by $r^4$ to facilitate a clearer comparison of the different cases: For
the case of $p=0$, the outer profile is,  as described in previous studies  \citep{joyce_etal_2009, syloslabini_2012,SylosLabini_CapuzzoDolcetta_2020}, characterized by
a radial decay with  $\rho(r) \sim r^{-4}$. Both the amplitude of the (flat, approximately constant $\rho$)
core at smaller scales and the amplitude of this tail are strongly $N$ dependent. 
As discussed in \cite{SylosLabini_CapuzzoDolcetta_2020}, the exponent $-4$ corresponds 
 to the limiting situation in  which the system is in a Jeans's equilibrium state; in this state the mass 
 is concentrated in the core, so that the system gravitational potential decays as $\Phi \sim r^{-1}$ 
and bound particles in the outskirts  move on purely radial orbits
(corresponding to an anisotropic velocity parameter $\beta \rightarrow 1$).
This configuration is reached only when the collapse is so violent  that 
the potential well of the core becomes very deep and, correspondingly,
as seen above,  a very significant fraction of mass and energy
is ejected,  leaving the bound particles with an energy much lower
than that of the initial state. If the collapse is much softer, however,
the mass distribution is less compact and the gravitational  potential decays more slowly 
than $ \sim r^{-1}$. The particles have less strongly radial  orbits as 
these are a result of the energy ejection around maximal contraction.   
Figure \ref{P_rho} shows behaviors that are also very consistent with these
findings for our ICs. As $p$ increases we see a very apparent transition
from an $r^{-4}$ to an $r^{-3}$ outer tail in the profile. Further, in line
with the previous plots we have analyzed, we see that there appears to be good evidence for 
convergence with $N$ as one
approaches $p=1$. At $p=0.6$, however, there is still a strong 
$N$ dependence in both the shape and amplitude of the profile, 
and it remains unclear what the functional dependence may be
as $N \rightarrow \infty$. 

\begin{figure*} 
\includegraphics[width=7in,angle=0]{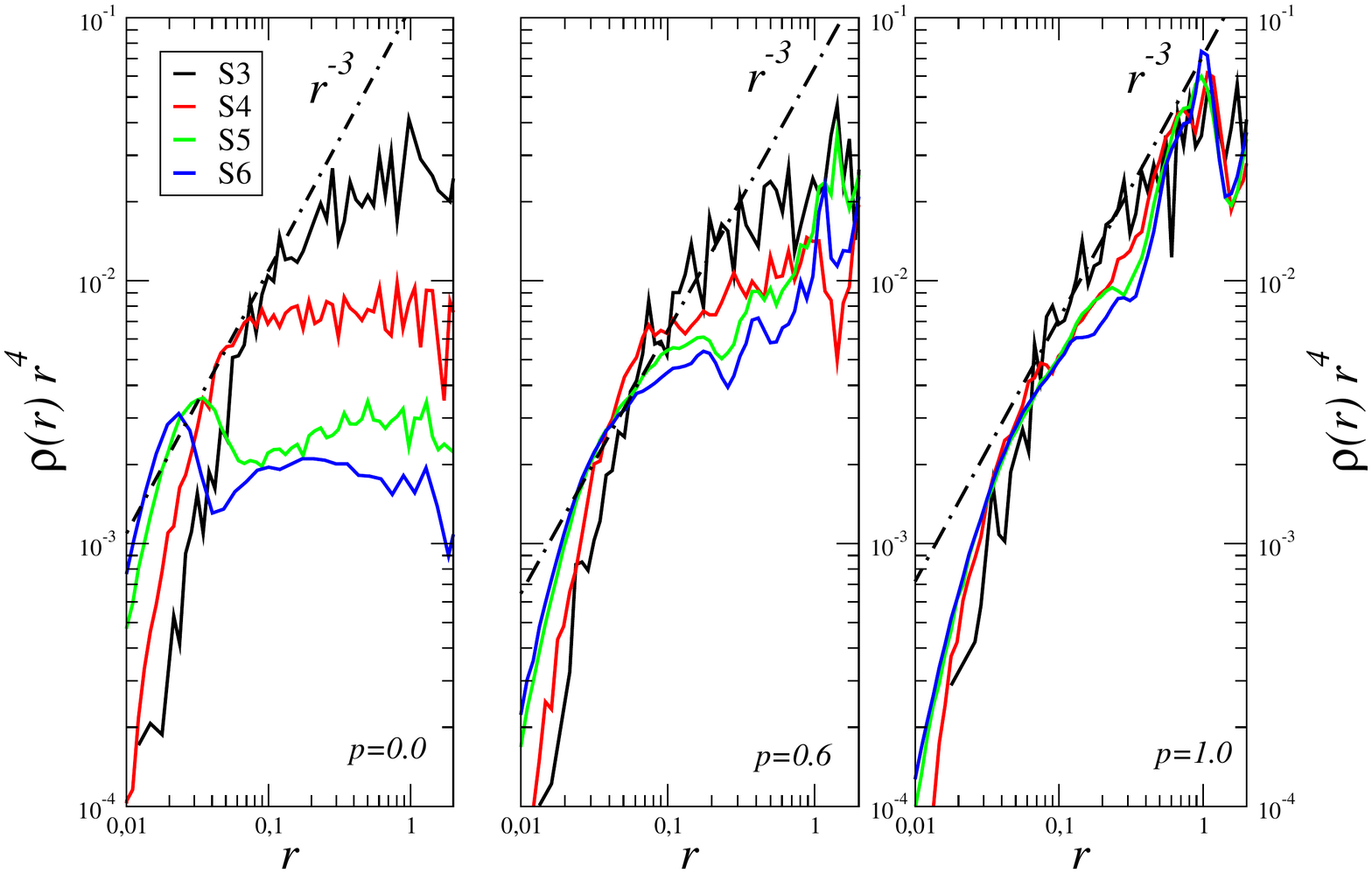}
\caption{Density profile multiplied by $r^4$ for $p=0$ (left panel), $p=0.6$ (middle panel), and $p=1$ (right panel)
obtained from Poisson configurations with different values of $N_0$. 
The softening length is   $\varepsilon = 3 \times 10^{-4}$.
The profile for $p=0$ is $N$ dependent  and decays as $r^{-4}$; for $p=1$  
it converges  to an $N$-independent $\approx r^{-3}$ behavior 
but is still affected by finite $N$ effects at small scales.
{A reference line corresponding to $\rho(r)\sim r^{-3}$
is reported in the middle and left panels.}}
\label{P_rho} 
\end{figure*}

Figure \ref{rho_glass} shows the measured density profile (multiplied by $r^4$) for  
glass ICs with $N_0=5 \times 10^4$ for different values of $p$. Its behavior
is qualitatively the same as what we saw above, with the same 
apparent transition in the functional form as a function of $p$.
We also find very similar 
behavior for the other quantities we have analyzed, for both the dependence on 
$p$ and on $N_0$. 
There is, however, a  quantitative difference between the Poisson and the glass ICs:
At given $p$, the glass ICs with a given $N_0$ closely resemble the Poisson ICs
at a slightly larger $N_0$.  For instance, a glass IC with $N_0 \approx 10^4$ 
results in a profile that agrees well with that of a Poissonian IC with $N_0 \approx 10^5$. 
This result is completely in line with what was anticipated: At a given $N_0$, the 
very uniform glass configuration has density fluctuations that are more suppressed compared 
to those of a Poisson configuration. Nevertheless, the quantitative effect, though measurable, is  
relatively small and our conclusions for what concerns convergence are not changed by
our study of these ICs.
\bef
\includegraphics[width=2.8in,angle=-90]{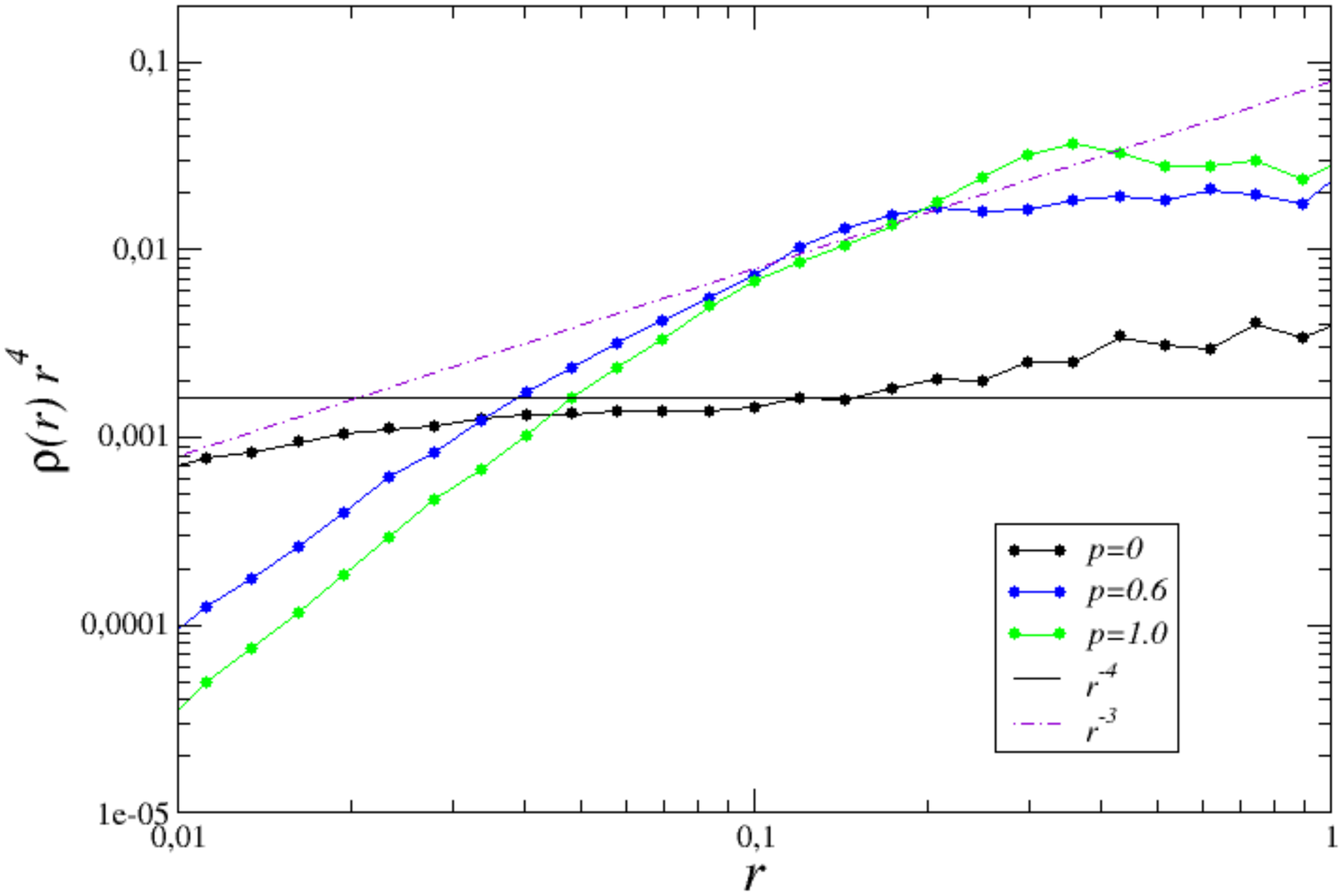}
\caption{Density profiles multiplied by $r^4$ for simulations 
with glass ICs and $N_0=5 \times 10^4$ for different values of $p$.
Two reference lines with exponents $-4$ and $-3$ are also plotted.
The softening length is $\varepsilon =  3 \times10^{-4}$.} 
\label{rho_glass}
\eef

\section{Conclusions}
\label{Conclusions}

We have numerically studied the collapse and virialization of isolated and initially cold configurations of particles 
with uniform mean density and an irregular asymmetric shape specified by a parameter, $p$, that characterizes the
deviation from sphericity. Such ICs can be considered as a simple schematic representation of the
first collapsing regions in cosmological models with a lower cutoff in the spectrum of initial fluctuations. 
In this context, the number of particles, $N$, is an unphysical parameter that arises from the $N$-body discretization,
and the model allows us to explore convergence to results that do not depend on it. More generally, the model
is a simple toy model for studying how the collapse at finite $N$ in the spherical limit may be modified by 
deviations from spherical symmetry.

Studying configurations with $p$ in a range where the collapse remains 
monolithic (i.e., well characterized as a 
single collapse around a center) and $N$ in the range $10^3$ to $10^6$,  we have been able to see an apparent 
transition in the qualitative nature of the collapse and the properties of the virialized  QSS that result from it.
This is most evidently shown by the outer density profile that changes from a power-law decay 
with an $r^{-4}$ behavior in the cold spherical limit to a decay with $r^{-3}$ when the rotational symmetry
is broken sufficiently strongly. There are likewise differences that reflect such a transition in the several other various macroscopic quantities we have considered to characterize both the dynamics during the
collapse and the final relaxed QSS. The $p$ and $N$ dependence of all these quantities leads to a very
coherent interpretation in terms of a competition between the density fluctuations associated with the
deviations from spherical symmetry and those associated with the discrete particle distribution.
As $p$ increases, the $N$ dependence progressively weakens, and at the largest $p$ there is
reasonable evidence for convergence to $N$-independent results. Nevertheless, residual $N$ 
dependences are still clearly detectable in some quantities -- in the inner parts of the density profile and in the
tails of the particle energy distribution -- even for the very largest $N$. 
In the case of an initially spherical cloud, the transition between  a density power-law decay 
with an $r^{-4}$ behavior  to a decay with $r^{-3}$ is regulated by density fluctuations  \citep{SylosLabini_CapuzzoDolcetta_2020}.

In relation to cosmological structure formation, our results qualitatively illustrate two important points.
Firstly, they show that a very simple breaking of spherical symmetry in a cold collapse appears to be sufficient 
to lead to the emergence of an $r^{-3}$ tail in the virialized state. 
Such a tail (in the density profiles of
halos, i.e., quasi-virialized clumps) is seen very generically in cosmological simulations and 
has been proposed as a ``universal''   that is independent of ICs 
(see, e.g., \citealt{Navarro_etal_1996}). Numerous studies 
of cold collapse with much more complex ICs than ours have supported such universality
\citep[see, e.g.,][]{Navarro_etal_2004,Wang_2020}, but its origin has remained elusive. Our simple ICs provide a very
simplified model for studying this fundamental issue:
The nature of the outer profiles of cold dark matter halos  is related to 
both the breaking of spherical symmetry in a cold collapse and the effect of the density fluctuations.

Secondly, our results indicate that the 
resolution (i.e., particle density)  needed to resolve the continuum limit of the first cold collapses 
predicted in many models  of cosmological  structure formation may be extremely difficult to reach. 
Indeed, we have seen that even for quite asymmetric ICs (e.g., corresponding to $p=0.6$
in our model) we do not see an approach to convergence of basic macroscopic 
quantities, even for $N=10^6$ particles. This is many times larger than the typical
number of particles inside the first collapsing structures in simulations 
of warm dark matter models  that attempt to model  such collapses \citep{Wang_White_2007,KUHLEN201250}. 
On the other hand, we have seen that the number of particles, $N$, needed 
to reach convergence is manifestly strongly dependent on the parameter $p$,
and thus we cannot draw quantitative conclusions for the resolution needed
in such cosmological simulations. In future work we plan to address 
this issue by trying to more precisely model how the breaking of spherical
symmetry at the onset of collapse can be characterized and determined
in such cosmological models.

\begin{acknowledgements}
We are grateful to Lehman Garrison for having provided us the glassy configuration. 
We thank Roberto Capuzzo Dolcetta, Andrea Gabrielli {and Pierfrancesco Di Cintio} for useful discussions.
{Finally, we also thank the referee    for his useful comments.}
\end{acknowledgements}

\end{document}